\documentclass[prl,reprint,superscriptaddress]{revtex4-1}

\pdfoutput=1

\usepackage{amsmath,amssymb, graphicx, hyperref, natbib}

\newcommand{\+}{^\dagger}
\newcommand{\nodag}{^{\phantom\dagger}}
\newcommand{\nostar}{^{\phantom*}}
\newcommand{\hH}{\hat{H}}
\newcommand{\hc}{\hat{c}}
\newcommand{\hd}{\hat{d}}
\newcommand{\ket}[1]{\left| #1 \right\rangle}
\newcommand{\bra}[1]{\left\langle #1 \right|}
\newcommand{\braket}[2]{\left\langle #1\middle| #2\right\rangle}
\newcommand{\expect}[1]{\left\langle #1 \right\rangle}
\newcommand{\vk}{\vec{k}}
\DeclareMathOperator{\sech}{sech}

\begin{document}
\author{Bhuvanesh Sundar}
\email{bs55@rice.edu}
\affiliation{Department of Physics and Astronomy, Rice University, Houston TX 77005, USA}
\affiliation{Rice Center for Quantum Materials, Rice University, Houston TX 77005, USA}

\author{Bryce Gadway}
\email{bgadway@illinois.edu}
\affiliation{Department of Physics, University of Illinois at Urbana Champaign, Urbana IL 61801-3080, USA}

\author{Kaden R. A. Hazzard}
\email{kaden@rice.edu}
\affiliation{Department of Physics and Astronomy, Rice University, Houston TX 77005, USA}
\affiliation{Rice Center for Quantum Materials, Rice University, Houston TX 77005, USA}

\title{Synthetic dimensions in ultracold polar molecules}
\date{\today}

\begin{abstract}
Synthetic dimensions alter one of the most fundamental properties in nature, the dimension of space. They allow, for example, a real three-dimensional system to act as effectively four-dimensional. Driven by such possibilities, synthetic dimensions have been engineered in ongoing experiments with ultracold matter. We show that rotational states of ultracold molecules can be used as synthetic dimensions extending to many -- potentially hundreds of -- synthetic lattice sites. Microwaves coupling rotational states drive fully controllable synthetic inter-site tunnelings, enabling, for example, topological band structures. Interactions leads to even richer behavior: when molecules are frozen in a real space lattice with uniform synthetic tunnelings, dipole interactions cause the molecules to aggregate to a narrow strip in the synthetic direction beyond a critical interaction strength, resulting in a quantum string or a membrane, with an emergent condensate that lives on this string or membrane. All these phases can be detected using measurements of rotational state populations.
\end{abstract}

\maketitle
\section{Introduction}
Ultracold polar molecules offer unique possibilities for creating strongly correlated matter, owing to their strong anisotropic long-ranged dipolar interactions and their complex rotational and vibrational structure~\cite{lemeshko2013manipulation, moses2017new, gadway2016strongly, carr2009cold, fedorov2016novel, sundar2013universal, brennen2007designing, micheli2006toolbox, gorshkov2011tunable, gorshkov2011quantum, manmana2013topological, manmana2017correlations, barnett2006quantum, wall2013simulating}. Although previous experimental and theoretical research has utilized the rotational degree of freedom~\cite{gorshkov2011tunable, gorshkov2011quantum, manmana2013topological, manmana2017correlations, wall2013simulating, gorshkov2013kitaev, hazzard2013far, hazzard2014many, yan2013observation, barnett2006quantum, wall2015realizing, wall2015quantum, glockner2015rotational}, it has used only a few rotational or dressed rotational states.

In this article, we propose to use rotational states of polar molecules as a synthetic dimension, which can have up to hundreds of synthetic lattice sites. The synthetic tunnelings are driven by microwaves resonant with rotational state transitions. This gives rise to a system with a fully tunable synthetic single particle Hamiltonian, which experiments can use to realize arbitrary synthetic band structures, including topological ones. We show that dipole interactions in polar molecules lead to interesting phases, even without any special engineering or fine tuning. For example, we show that molecules frozen in a periodic real space array undergo a spontaneous dimensional reduction, forming a fluctuating quantum string or membrane. At strong interactions, the string/membrane hosts an emergent condensate of hardcore bosons. We show that ongoing experiments can realize and probe these strings/membranes and condensate.
\begin{figure}[t]
\includegraphics[width=1.0\columnwidth]{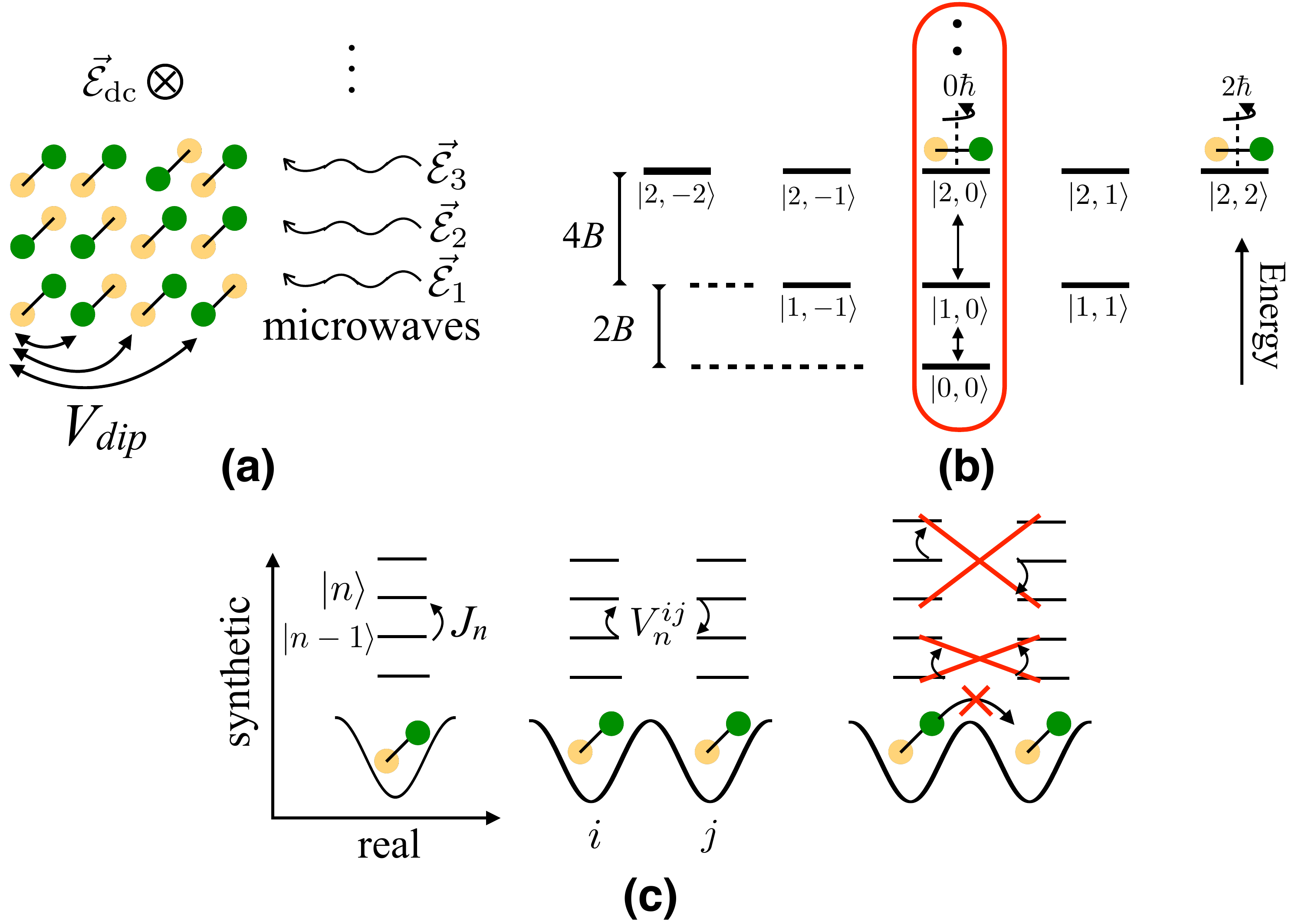}
\caption{\label{fig: setup}(a) Schematic of setup. Molecules in a periodic (1D or 2D) array in the $x$-$y$ plane are driven by microwaves, and interact via long-ranged dipole interactions. (b) A synthetic dimension is formed by the rotational states circled in red, in addition to the real spatial dimensions. Vertical arrows indicate transitions driven by microwaves (see also (c)). Here, $\ket{n,m}$ refers to a rotational state with total and azimuthal angular momentum $n$ and $m$. (c) Two types of processes occur in this system: Microwaves $\vec{\mathcal{E}}_n$ drive effective tunneling $J_n \propto \left|\vec{\mathcal{E}}_n\right|$ (left), and correlated tunneling $V_n^{ij}$ arise from dipole interactions between molecules at real lattice sites $i$ and $j$ (middle). Off-resonant processes (right) disappear in the rotating-wave approximation. Dipole-induced transitions to states outside the circled set in (b) are also made off-resonant by a static electric field.}
\end{figure}

Researchers have created nearly quantum degenerate gases of several heteronuclear molecular species, such as KRb, NaRb, NaK, and RbCs, in their ground state~\cite{ni2008high, takekoshi2014ultracold, molony2014creation, park2015ultracold, guo2016creation}. All of these have a strong electric dipole moment of about a Debye. These molecules also have a large number of rotational quantum states. We define a synthetic lattice, whose sites are a subset of a molecule's rotational states. To create a large synthetic lattice, we propose to shine several microwaves in parallel to drive transitions up to a highly excited rotational state, as illustrated in Fig.~\ref{fig: setup}. These transitions correspond to tunneling in the synthetic lattice. Experimentalists can simultaneously apply a large number of microwaves with fully controllable amplitudes, phases, and frequencies ranging from a few to several tens of GHz using commercially available technology (see Supplementary Materials).

Experimentalists have created synthetic dimensions in other ultracold gases from their motional~\cite{an2017ballistic, an2017correlated}, spin~\cite{mancini2015observation, stuhl2015visualizing, celi2014synthetic, anisimovas2016semisynthetic}, clock~\cite{wall2016synthetic, livi2016synthetic, kolkowitz2017spin}, or rotational~\cite{floss2015observation} states that are coupled by Raman lasers, analogous to our proposal's coupling of rotational states with microwaves. Our proposal shares some features with these other methods. We can fully control every tunneling amplitude and on-site potential by tuning the microwaves' complex amplitudes and detunings. By appropriate choice of the rotational states, we can impose periodic or open boundaries on the synthetic lattice, or create other spatial topologies. We can image populations in the synthetic lattice with single-site resolution.

Additionally, realizing synthetic dimensions in polar molecules has significant advantages over other systems. First, the experimentally feasible size of the synthetic dimension is orders of magnitude larger. Second, since the internal states are directly coupled via microwaves without an intermediate excited state, the system does not suffer from heating encountered in schemes that employ two-photon Raman processes. Third, our system is insensitive to magnetic field noise that limits other methods. Finally, strong dipole interactions lead to rich many-body physics at a favorable energy scale.

\subsection{Setup.}
We consider a unit-filled periodic array of molecules trapped in the $x$-$y$ plane in an optical lattice or a microtrap array~\cite{lester2015rapid, kaufman2014two, endres2016atom, liu2017ultracold}, as illustrated in Fig.~\ref{fig: setup}. Current experiments achieve $\sim25\%$ filling~\cite{moses2015creation}, and experimental advances are steadily increasing this number. We impose a sufficiently deep lattice to completely suppress tunneling in real space. This avoids problematic molecular reactions~\cite{yan2013observation, chotia2012long, zhu2014suppressing, ospelkaus2010quantum, ni2010dipolar, de2011controlling} or complicated collision processes~\cite{mayle2012statistical, mayle2013scattering, doccaj2016ultracold, wall2017microscopic, wall2017lattice, ewart2017bosonic} that occur if two molecules occupy a single lattice site.
\begin{figure}[t]
\includegraphics[width=1.0\columnwidth]{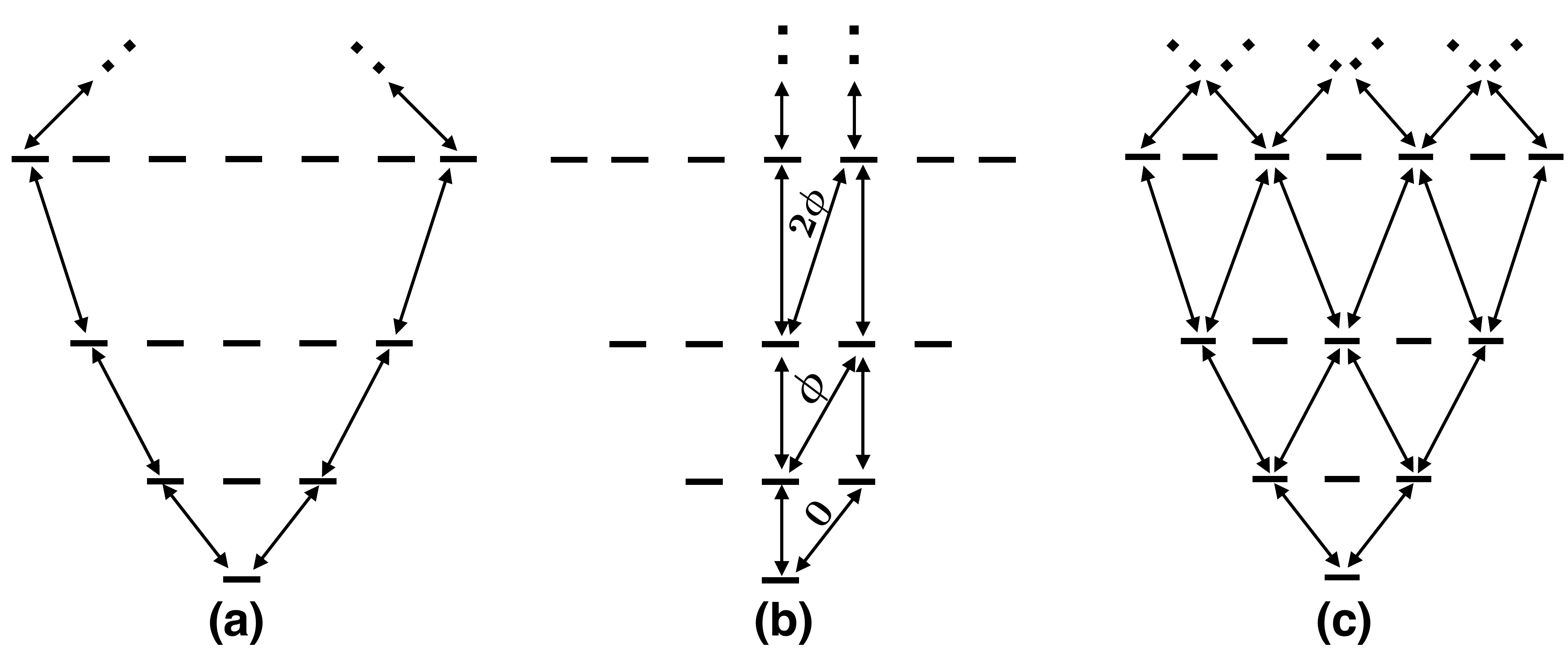}
\caption{\label{fig: top.}Three examples of engineering a synthetic dimension(s) using the internal rotational states of a molecule, each resulting in a different single-particle Hamiltonian. (a) 1D chain with periodic boundaries. (b) Two-leg ladder with complex tunnelings. The phase of the tunneling is indicated on the ladder's rungs. (c) Square lattice with open boundaries.}
\end{figure}

To create a 1D synthetic lattice with $N_{\rm rot}$ sites and open boundaries, the molecules are driven by $N_{\rm rot}-1$ microwaves, as shown in Fig.~\ref{fig: setup}. The polarization of the microwaves is chosen to yield the desired sign for the amplitude of angular momentum exchange driven by the dipole interaction. As we will show, microwaves that are linearly polarized in the $z$ direction yield a positive sign, while microwaves with a left-circular polarization with respect to the $z$ axis yield a negative sign. For linearly polarized microwaves, the synthetic lattice sites lie in the $\ket{n,0}$ subspace of the single-molecule rotational eigenstates, with $\hat{z}$ the quantization axis. In this case, the $n^{\rm th}$ microwave is resonant with the transition from $\ket{n-1,0}$ to $\ket{n,0}$. We apply a small electric field to detune the $\ket{n,m\neq0}$ states, so that molecules remain in the $\ket{n,0}$ space. The detuning due to the electric field is larger than dipole interactions and hyperfine mixing even for moderate electric fields $\sim\mathcal{O}(10)$~V/cm~\cite{will2016coherent}. For left-circularly polarized microwaves, the synthetic lattice sites lie in the $\ket{n,n}$ subspace, and a small electric field detunes away the $\ket{n,m\neq n}$ subspace. We discuss the technical details for applying the microwaves and accuracy of the resulting effective Hamiltonian in more detail in the Supplementary Material.

In the rotating wave approximation, our system is described by
\begin{equation}
\hH = -\sum_{nj} J_n \hc_{n-1,j}\+\hc_{nj}\nodag + \sum_{nij} V_n^{ij} \hc_{n-1,i}\+\hc_{ni}\nodag\hc_{nj}\+\hc_{n-1,j}\nodag + {\rm h.c.},
\label{eqn: H}
\end{equation}
where $\hc_{nj}\nodag$ ($\hc_{nj}\+$) annihilates (creates) a molecule on the real lattice site $j$ and synthetic lattice site $n$. Here, $J_n$ is the synthetic tunneling amplitude induced by a resonant microwave, and $V_n^{ij}$ is the angular momentum exchange amplitude induced by dipole interaction. When the synthetic lattice is the $\ket{n,0}$ subspace, $J_n = d\mathcal{E}_n^{(0)}\frac{n}{\sqrt{4n^2-1}}$, and $V_n^{ij} = \frac{4n^2}{4n^2-1}\frac{Va^3}{r_{ij}^3}$, with $\mathcal{E}_n^{(0)}$ the $n^{\rm th}$ microwave's amplitude, $d$ the permanent electric dipole moment, $a$ the real lattice constant, and $V=\frac{d^2}{32\pi\epsilon_0a^3}$ [see Supplementary material]. When the synthetic lattice is the $\ket{n,n}$ subspace, $J_n = d\mathcal{E}_n^{(0)}\frac{n}{\sqrt{2n+1}}$ and $V_n^{ij} = \frac{2n}{2n+1}\frac{Va^3}{r_{ij}^3}$, with $V = \frac{-d^2}{16\pi\epsilon_0a^3}$. For both choices of synthetic lattice sites, $V_n^{ij}\approx V\frac{a^3}{r_{ij}^3}$ is nearly independent of $n$ for large $n$. Dipole-induced processes that change the total azimuthal angular momentum are off-resonant and average to zero in the rotating frame. Each tunneling amplitude $J_n$ can be tuned by adjusting the amplitude and phase of $\mathcal{E}_n^{(0)}$. Although set to zero in Eq.~\eqref{eqn: H}, an on-site potential can be introduced on the $n^{\rm th}$ synthetic lattice site by detuning the $(n-1)^{\rm th}$ and $n^{\rm th}$ microwaves. We note that since the molecules are stationary, it is irrelevant whether $\hc_{nj}$ are fermionic or bosonic operators. Without loss of generality, we assume they are fermionic.

The above setup implements a 1D synthetic lattice with open boundaries. Using a more sophisticated microwave and static field architecture, it is also possible to create arbitrary synthetic lattices, for example a synthetic lattice with periodic boundaries as in Fig.~\ref{fig: top.}(a), a synthetic lattice under a gauge field as in Fig.~\ref{fig: top.}(b), two synthetic dimensions as in Fig.~\ref{fig: top.}(c), or even other topologies. One can also engineer higher lattice connectivities through higher order (e.g. two-photon or three-photon) microwave transitions. We emphasize that this ability to produce a fully controllable sophisticated single particle Hamiltonian is a key advantage of using molecules over atoms.

In this article, we consider the many-body physics in the simplest case of one open-boundary synthetic dimension with uniform and positive $J_n$, and show that dipole interactions lead to rich physics. We assume that the real space lattice is a one-dimensional (1D) chain or a two-dimensional (2D) square lattice. For the square lattice, for the microwave polarizations in our setup, the sign of $V_n^{ij}$ is isotropic in the real lattice plane. We first determine the two-molecule ground state, then examine the many-body behavior.

\section{Results}
\subsection{Bound state of two molecules.}
\begin{figure}[t]
\includegraphics[width=1.0\columnwidth]{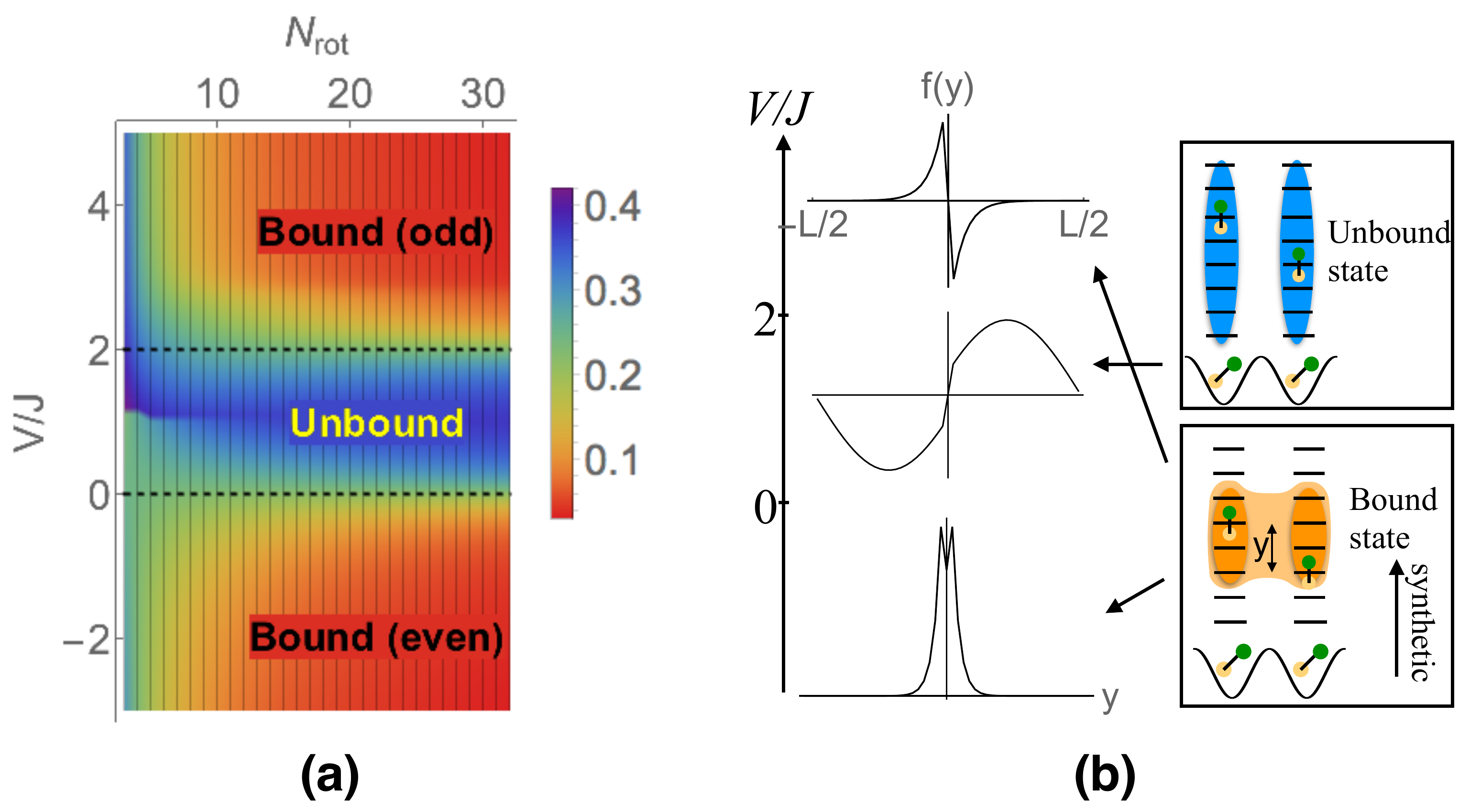}
\caption{\label{fig: phaseDiag 2mols}Two-molecule phase diagram. (a) Normalized average distance $\delta$ between the two molecules in the synthetic direction, as a function of interaction strength $V/J$ and synthetic dimension size $N_{\rm rot}$. The ground state undergoes binding transitions at $V=0$ and $V=2J$ (dotted lines) in the thermodynamic limit. (b) Representative relative wavefunctions in the three phases.}
\end{figure}
We exactly solve Eq.~\eqref{eqn: H} for two molecules. The solution already gives considerable insight into understanding the many-body phase diagram. We analytically solve the problem in the case that $V_n=V$ are uniform, and $N_{\rm rot}$ is large. We also perform a numerical calculation for finite $N_{\rm rot}$ and physical values of $V_n$, and find that the uniform $V_n$ limit captures the essential physics.

Writing the most general state for two molecules, $\ket{\psi} = \sum_{mn} f_{mn} \hc_{m1}\+\hc_{n2}\+\ket{\rm vac}$, we find that there are three types of solutions for $f_{mn}$ in three different regions of parameter space. Figure~\ref{fig: phaseDiag 2mols}(b) shows the behavior of the ground state in these three regions. For $0<V<2J$, the two molecules are in a scattering state, $f_{mn} = e^{\pm i(m-n)}$ for $m\neq n$, with a scattering phase shift as the molecules cross each other in the synthetic dimension. In this regime, the molecules are delocalized throughout the synthetic direction. If $V/J<0$ or $V/J>2$, the molecules are localized in a bound state, $f_{mn} = e^{-\lambda(m-n)}$ for $m>n$, with a binding length $1/\lambda$. The bound state is even under exchange of the two molecules if $V/J<0$, and odd if $V/J>2$. The binding length diverges at the critical points $V/J=0$ and $2$, and it monotonically decreases away from these points. As $V/J\rightarrow\pm\infty$, the two molecules are tightly bound in a state that is two synthetic lattice sites wide:
\begin{equation}
\ket{\psi_n} = \frac{ \hc_{n,1}\+\hc_{n+1,2}\+ \pm \hc_{n+1,1}\+\hc_{n,2}\+ }{\sqrt{2}} \ket{\rm vac},
\label{eqn: psi2mol}
\end{equation}
where $n$ is arbitrary.

The same essential physics exists for finite $N_{\rm rot}$ and nonuniform physical couplings $V_n = V\frac{4n^2}{4n^2-1}$. We demonstrate this by numerically diagonalizing Eq.~\eqref{eqn: H}. We characterize the ground state by the normalized relative molecular separation in the synthetic direction, $\delta = \frac{1}{N_{\rm rot}}\sum_{mn} |m-n|\expect{\hc_{m1}\+\hc_{m1}\nodag\hc_{n2}\+\hc_{n2}\nodag}$, which we plot in Fig.~\ref{fig: phaseDiag 2mols}(a). For $N_{\rm rot}\rightarrow\infty$, $\delta$ asymptotes to $\frac{e^\lambda}{2N_{\rm rot}\sinh\lambda}$ in the bound states, and undergoes a sharp transition at $V=0$ and $V=2J$. For finite $N_{\rm rot}$, the transition is smoother, and occurs at larger $|V/J|$.

We emphasize that the ground states of our system differ from conventional magnetically ordered/disordered states of a large-spin system, with the synthetic lattice sites mapped to spin states. In our system, the amplitudes $J_n$ and $V_n$ (for large $n$) are uniform, leading to a translational symmetry in the synthetic direction. This is a highly unnatural Hamiltonian for a large-spin system, and is reflected in the structure of the ground state. The relative width of the ground state in the synthetic direction is finite, unlike ferromagnetic phases that occupy width $\mathcal{O}\left(\sqrt{N_{\rm rot}}\right)$.

\subsection{Many-body phase diagram.}
\begin{figure}[t]
\includegraphics[width=1.0\columnwidth]{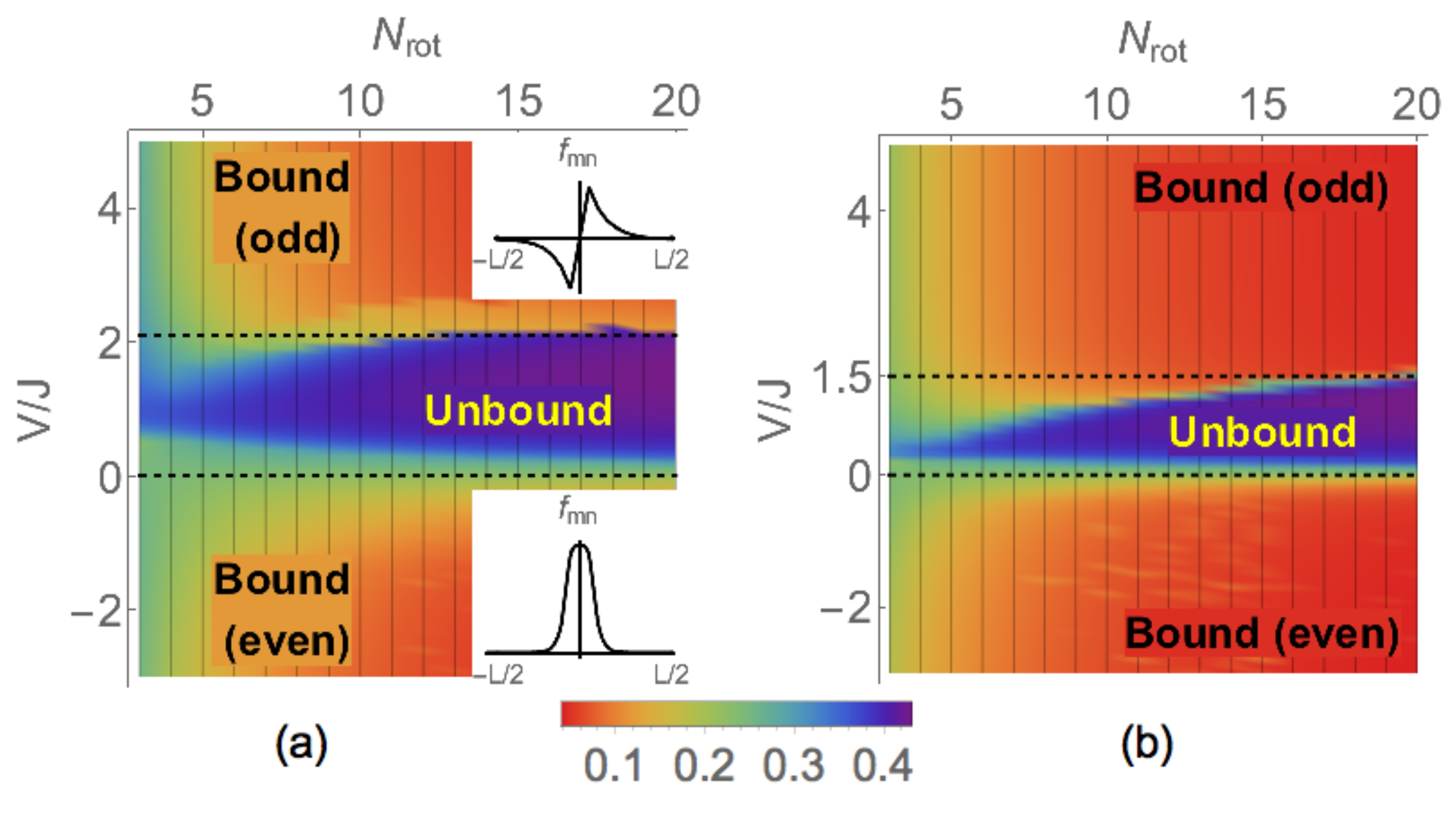}
\caption{\label{fig: phaseDiag}Many-body phase diagram for a) a one-dimensional chain, and b) a square lattice in real space. When $V\lesssim0$ or $V\gtrsim2.15J$ [dotted lines in (a)] in a 1D chain of molecules, adjacent molecules bind in the synthetic dimension. Bound pairs of molecules align and collapse to a quantum string. The color scale denotes the normalized average synthetic distance $\delta$ between molecules in a bound pair. For a 2D array, the molecules collapse to a membrane when $V\lesssim0$ or $V\gtrsim1.5J$ [dotted lines in (b)]. In the intermediate regime, the molecules form a gas. Insets: Representative variational wavefunctions in the quantum string phases.}
\end{figure}
Next we explore Eq.~\eqref{eqn: H} for 1D and 2D periodic real space arrays of molecules. For a 1D array, we assume a variational ansatz that reproduces the exact solution for two molecules:
\begin{equation}
\ket{\psi_{\rm var}} = \prod_{i\in{\rm even}}\sum_{mn} f_{mn} \hc_{mi}\+\hc_{n,i+1}\+ \ket{\rm vac}. \label{eqn: psivar}
\end{equation}
This ansatz can also be viewed as a ``cluster mean field'' approximation~\cite{fazekas1991cluster} on which the system is divided into pairs of sites in real space that are coupled through the mean field.

Figure~\ref{fig: phaseDiag}(a) shows the phase diagram found by minimizing the energy with respect to the variational parameters $f_{mn}$. There are three phases of matter, each corresponding to a type of two-body state found above. In the bound phases, pairs of adjacent molecules bind together in the synthetic direction. Adjacent bound pairs similarly attract each other in this direction. As a result, the system spontaneously collapses to a one-dimensional string. The width of the string in the synthetic dimension varies with $V/J$, from two sites wide at $V/J\rightarrow \pm\infty$, to a diverging value at the transitions. When $N_{\rm rot}\gg1$, the transitions occur at $V=0$ and $V=2.15J$. The molecules are unbound when $0<V<2.15J$. The transitions are smoothed out and shifted at finite $N_{\rm rot}$.

When $V/J=\pm\infty$, the system spontaneously breaks the synthetic translational symmetry and a $U(1)$ symmetry of the Hamiltonian. Each molecule along the quantum string spontaneously localizes to only two states $\ket{n}$ and $\ket{n+1}$, forming a string that is two synthetic sites wide, breaking the synthetic translational symmetry. The system can be effectively thought to host hardcore bosons, with $\ket{n}$ corresponding to a vacant site and $\ket{n+1}$ to a singly occupied site. In this description, dipole interactions look like tunneling for the hardcore bosons between real lattice sites, leading to a hardcore Bose-Einstein condensate living on the string as the ground state, with a corresponding broken $U(1)$ symmetry.

We also consider a 2D array of molecules in the $x$-$y$ plane. We extend the mean field ansatz in Eq.~\eqref{eqn: psivar} to 2D, and variationally minimize the energy to calculate the phase diagram, which we plot in Fig.~\ref{fig: phaseDiag}(b). We again find binding transitions at $V\lesssim0$ and $V\gtrsim1.5J$, beyond which the molecules form a quantum membrane. We again find condensate transitions at $V/J=\pm\infty$.

To assess the accuracy of our approximation, we investigate two other variational ansatzes: a single-site mean field, and a mean field theory of fermionic pairs. We find that the phase diagram in all cases is similar to the cluster mean field results. The broad concurrence of these results from very different approximations gives us confidence in the basic physics of our model. In all cases, there are three phases: two phases with molecules forming bound states, and one of unbound molecules. Only the details of the long-range correlations differ among the approximations. We describe our other variational ansatzes in detail in the Supplementary Material.

Although earlier works~\cite{yin2011stable, fu2012separation, capogrosso2011superfluidity, wang2006quantum} have obtained broadly similar bound states in models without synthetic dimensions, the situation presented here has experimental advantages and differs in some of the relevant physics.
By utilizing a synthetic dimension, string states appear without needing molecules to tunnel in real space, thus avoiding the chemical reactions~\cite{yan2013observation, chotia2012long, zhu2014suppressing, ospelkaus2010quantum, ni2010dipolar, de2011controlling} and complex collisional processes~\cite{mayle2012statistical, mayle2013scattering, doccaj2016ultracold, wall2017microscopic, wall2017lattice, ewart2017bosonic} that will occur in other proposals and almost surely lead to significant loss or heating. Additionally, the energy scales are more favorable in our proposal. While earlier proposals required the motional degrees of freedom to be cooled below the dipole interaction scale to observe strings, our proposal only requires the rotational state temperature to be below this scale. Experiments routinely achieve near-zero entropy rotational state superpositions, which satisfies our requirement. The phases in the present system also somewhat differ from earlier proposals. First, a condensate lives on them, and, second, we show that one can obtain membranes, which have not appeared in previous work.

\section{Discussion}
\subsection{Experimental detection.}
The phases in our system can be detected by unitary transformations of rotational states and measurements of ground state populations. Populations in all rotational states can be selectively measured by a direct absorptive image~\cite{wang2010direct} or a time-resolved image taken with a resonantly enhanced multiphoton ionization~\cite{deiglmayr2008formation}. The binding between the molecules can be detected by measuring the distribution of molecules in the synthetic dimension. In the bound phase, adjacent molecules are relatively close to each other in the synthetic dimension. Specifically the relative spread $\delta$ of molecules in the synthetic dimension is $\mathcal{O}(1)$. In the unbound phase, $\delta\sim\mathcal{O}(N_{\rm rot})$ is large. The even/odd parity of the bound state and the $U(1)$ symmetry breaking can be characterized by the single-particle coherences between rotational states, which can be measured by performing $\pi/2$ unitary transformations between the desired rotational states before measuring populations.

\subsection{Variations of the experimental setup.}
The full control over the single particle Hamiltonian in our system results in the potential to explore a wide variety of other physics. For example, staggering the amplitudes $J_n$ would lead to a topological band structure~\cite{su1979solitons, heeger1988solitons, hasan2010colloquium, qi2011topological}. Randomizing the microwave detunings -- which we set to zero in Eq.~\eqref{eqn: H} -- or the tunneling amplitudes would allow us to explore physics related to Anderson and many-body localization~\cite{anderson1958absence, an2017ballistic}. Periodic modulation of the microwave amplitudes or frequencies will allow us to realize effective interactions between three or more particles in the Floquet picture, which will result in novel physics~\cite{rapp2012ultracold}. Engineering a two-dimensional synthetic lattice with complex tunneling amplitudes would implement synthetic gauge fields, manifesting for instance the Hofstadter butterfly~\cite{peierls1997theory, hofstadter1976energy}. Reintroducing tunneling between real lattice sites also allows access to a rich variety of physics, extending the $N_{\rm rot}=2$ case studied in Refs.~\cite{gorshkov2011quantum, gorshkov2011tunable, manmana2017correlations}. In the presence of real-space tunneling, multiple molecules can occupy the same lattice site in the lowest band, and dipole interaction produces a synthetic nearest-neighbor interaction between molecules that lie on the same real lattice site. These nearest-neighbor interactions lead to charge density wave or $p$-wave superfluid order in the synthetic direction~\cite{hirsch1985attractive, kuhner2000one, kitaev2001unpaired}.

\section{Summary}
We have shown that the rotational states of polar molecules can act as a highly controllable synthetic dimension. The tunnelings and band structure in this dimension can be precisely tuned by arbitrary microwave waveforms, and probed by established spectroscopic tools. This opens avenues to explore physics related to Anderson and many-body localization, synthetic gauge fields, topological band structures, and topological superfluids. The unusual form of dipole interactions in the synthetic dimension opens up new types of phenomena to ultracold matter.

We examined the many-body physics arising from interactions between polar molecules, in the simplest experimental scenario where microwaves drive tunneling of molecules in a synthetic dimension with a uniform amplitude, and where real space tunneling is suppressed. We found intriguing forms of correlated quantum matter, including fluctuating quantum strings and membranes, on whose surfaces live strongly interacting condensates. We emphasize that this physics emerges naturally with no intricate engineering or fine tuning, and the energy scales are large -- set directly by the dipole interaction. Moreover, because our setup avoids double occupancies in the real space lattice, the experimental lifetime is much longer than the $\sim$ms timescales associated with the dipole interaction. It will be fascinating for future research to explore these membranes, the interplay between the quantum fluctuations of the membrane and the condensate that lives on it, and their stability in the presence of perturbations such as disorder.

\bibliography{BibFile}

\renewcommand{\theequation}{S\arabic{equation}}
\renewcommand{\thefigure}{S\arabic{figure}}
\setcounter{equation}{0}
\setcounter{figure}{0}

\begin{widetext}
\section{Supplementary Materials for \textit{Synthetic dimensions in ultracold polar molecules}}
\section{Effective Hamiltonian}
In the lab frame, our system is described by the Hamiltonian
\begin{equation}
 \hH = \sum_j B\hat{N}_j^2 + \sum_j\sum_{\substack{n=1}}^{N_{\rm rot}}\hat{\vec{d}}_j\cdot\vec{\mathcal{E}}_n(t) + \sum_{ij} \frac{ \hat{\vec{d}}_i\cdot\hat{\vec{d}}_j - 3\left(\hat{\vec{d}}_i\cdot\hat{r}_{ij}\right)\ \left(\hat{\vec{d}}_j\cdot\hat{r}_{ij}\right) }{4\pi\epsilon_0r_{ij}^3},
\label{eqn: Hsupp}
\end{equation}
where $\hbar\hat{N}_j$ is the rotational angular momentum operator for the $j^{\rm th}$ molecule, and $B\sim\hbar\times\mathcal{O}\left({\rm GHz}\right)$ is the rotational constant. The eigenstates $\ket{n,m}$ of the first term in Eq. (\ref{eqn: Hsupp}) are described by two quantum numbers: the angular momentum quantum number $n$ and its projection $m$ on the $z$ axis. Equation~\eqref{eqn: Hsupp} assumes that the molecules are rigid rotors, i.e all molecules are in the lowest vibrational state, and coupling to excited vibrational states is small. This assumption is valid up to $n$ that approximately lies in a range between $40$ and $60$ in a typical diatomic molecule. (For example, for RbCs whose rotational constant is $\sim h\times500$~MHz and vibrational excitation energy is $\sim h\times1.5$~THz, the rotational excitation above the ground state is smaller than the vibrational excitation for $n\lesssim55$. For NaRb, whose rotational constant and vibrational excitation energy are $\sim h\times2$~GHz and $\sim h\times3.2$~THz, the rotational excitation above the ground state is smaller than the vibrational excitation for $n\lesssim40$). For larger $n$, the rotational states get dressed by rovibrational coupling.

The matrix elements for the dipole moment operator between rotational states in a rigid rotor are~\cite{brown2003rotational, gorshkov2011quantum}
\begin{align}
\bra{n,m}\hd^p\ket{n',m'} = & (-1)^pd\sqrt{\frac{4\pi}{3}} \int\int Y_{n'm'}(\Theta,\Phi)Y^*_{nm}(\Theta,\Phi)Y_{1p}\nostar(\Theta,\Phi) \sin\Theta d\Theta d\Phi\nonumber
\\=& (-1)^pd \sqrt{\frac{2n+1}{2n'+1}} \braket{n,m; 1,p}{n',m'} \braket{n,0;1,0}{n',0}, \label{eqn: dp}
\end{align}
where $\hd^p$ are the spherical components of the dipole operator, $d$ is the molecule's permanent electric dipole moment, and $p$ takes the values $0,\pm1$. Using the properties of Clebsch-Gordan coefficients, Eq. (\ref{eqn: dp}) simplifies to
\begin{align}
\bra{n,m}\hd^0\ket{n-1,m'} &= d\delta_{mm'}\sqrt{\frac{n^2-m^2}{4n^2-1}} \nonumber\\
\bra{n,m}\hd^\pm\ket{n-1,m'} &= \pm d\delta_{m,m'\pm1}\sqrt{\frac{(n\pm m)(n\pm m-1)}{2(4n^2-1)}} \nonumber\\
\bra{n,m}\hd^p\ket{n',m'} &= 0\ {\rm if}\ |n-n'|\neq1.
\label{eqn: d}
\end{align}
For larger $n$ where the rigid rotor assumption breaks down, the polarization of the molecules typically reduces due to a larger internuclear separation, so the matrix elements are smaller than Eq.~\eqref{eqn: d}. In this limit, matrix elements can be obtained from techniques in earlier works~\cite{kotochigova2003abinitio} which calculate dipole moments as a function of the internuclear separation.

We find from Eqs.~\eqref{eqn: Hsupp} and ~\eqref{eqn: d} that when the molecules are illuminated by a resonant $\hat{z}$-polarized microwave of amplitude $\mathcal{E}_n^{(0)}$, the synthetic tunneling strength from $\ket{n-1,0}$ to $\ket{n,0}$ is $J_n = d\mathcal{E}_n^{(0)}\frac{n}{\sqrt{4n^2-1}}$. When the molecules are illuminated by a $\sigma^+$-polarized microwave, the tunneling strength from $\ket{n-1,n-1}$ to $\ket{n,n}$ is $J_n = d\mathcal{E}_n^{(0)}\sqrt{\frac{n}{2n+1}}$. We choose the microwave fields such that $J_n=J$ is uniform in $n$: $\mathcal{E}_n^{(0)} = \frac{J\sqrt{4n^2-1}}{nd}$ for the $m=0$ states, and $\mathcal{E}_n^{(0)} = \frac{J\sqrt{2n+1}}{\sqrt{n}d}$ for $m=n$. Even when the dipole matrix elements are smaller than Eq.~\eqref{eqn: d} due to rovibrational mixing, the synthetic tunnelings can still be made uniform by compensating with stronger microwaves. To tune this, experiments can perform Rabi spectroscopy on the relevant two level system, and adjust the microwave amplitude until the desired tunneling rate is achieved.

Dipole interactions induce an angular momentum exchange with an amplitude $V_n^{ij}$, between two molecules in adjacent rotational manifolds. When the two molecules are in $m=0$ states, 
$V_n^{ij} = \bra{n-1,0}_i\bra{n,0}_j\hH_{\rm dd}\ket{n,0}_i\ket{n-1,0}_j = \frac{1-3(\hat{r}_{ij}\cdot\hat{z})^2}{8\pi\epsilon_0r_{ij}^3}\frac{d^2n^2}{4n^2-1}$, where $\hH_{\rm dd}$ is the third term in Eq.~\eqref{eqn: Hsupp}. For the $m=n$ states, $V_n^{ij} = \frac{3(\hat{r}_{ij}\cdot\hat{z})^2-1}{8\pi\epsilon_0r_{ij}^3}\frac{d^2n}{2n+1}$. These expressions are modified at large $n$, when rovibrational mixing is relevant~\cite{kotochigova2003abinitio}.

Interactions between the rotational and nuclear spin angular momenta introduce some mixing between the molecule's rotational and hyperfine spin quantum numbers. However, because we add a small electric field $\sim\mathcal{O}(10)$~V/cm, the energies of the different $m$ states within a rotational manifold split by $\sim10$-$100$~MHz. This splitting is typically larger than the interaction between nuclear and rotational angular momentum. As a result, the rotational angular momentum decouples from the nuclear spins, so $m$ is a good quantum number~\cite{will2016coherent}. The expressions for the matrix elements for the dipole moment operator [Eq.~\eqref{eqn: d}] are valid in this limit. Additionally, the splitting of the $m$ states within a rotational manifold is also larger than $V_n^{ij}$. Therefore, all dipole interaction-induced transitions to rotational states outside the Hilbert space in Eq.~\eqref{eqn: H} are off-resonant and suppressed. The electric field does mix states between different rotational manifolds $n$, but the resulting additions to the Hamiltonian in Eq.~\eqref{eqn: H} are insignificant for these electric field values.

\section{Microwave implementation}
Here, we briefly discuss the implementation of microwaves for the realization of the Hamiltonian in Eq.~\eqref{eqn: H}, and associated challenges. As a concrete example, we consider RbCs, whose rotational constant is $B \approx h\times500$~MHz. Cold and dense samples of RbCs have been produced by both the Innsbruck~\cite{takekoshi2014ultracold, reichsollner2017quantum} and Durham~\cite{molony2014creation} groups. The microwave frequency required to excite a molecule from the $\ket{n-1}$ to the $\ket{n}$ rotational manifold is $\omega_n = 2nB/\hbar \approx 2\pi n$~GHz. We consider the scenario where only the $m = 0$ rotational sublevels are populated, such that the number of synthetic lattice sites will be equal to the number of coupled rotational manifolds. In this case, an implementation of a 1D synthetic lattice with $N_{\rm rot}$ synthetic lattice sites and open boundary conditions requires $N_{\rm rot}-1$ microwaves, spanning (sparsely) the range from 1~GHz to $(N_{\rm rot}-1)$~GHz.

The generation, amplification, and projection of microwaves to coherently couple a large number of synthetic sites can be straightforwardly accomplished using standard and commercially available microwave equipment. The required microwave waveform itself, $\mathcal{E}(t) = \sum_{n=1}^{N_{\rm rot}-1} \mathcal{E}_n^{(0)}\cos(\omega_n t)$, can be generated with lab-developed or commercially available arbitrary waveform generators presently capable of reaching up to at least 20 GHz~\cite{raftery2017direct}. Additionally, since only a finite number of discrete frequencies are actually required from the large frequency span, versatile frequency sources generated by direct digital synthesis may be used to stroboscopically sample the various $\omega_n$. Standard broadband amplifiers and waveguide horns can then be used to project these signals onto the molecular samples.

At least for some molecules like RbCs, dozens of coupled synthetic sites should be straightforward to realize with present microwave technology. The number of coupled synthetic sites is fewer for molecules which have a larger rotational constant; for example, the synthetic lattice will contain roughly ten sites for NaRb or KRb. In addition to being limited by the frequency range of waveform generators, the number of coupled synthetic sites is also limited by the lifetime of excited rotational states. The decay rate of a molecule with typical parameters ($B=h\times500$~MHz and $d=1$D), from $n=20$ due to stimulated emission by a microwave field at room temperature, is $\sim1/$~sec. The decay rate is larger at larger $n$. Therefore for decay to be insignificant on an experimental timescale$\sim$~s, the number of coupled synthetic sites in this molecule is $n\lesssim20$. Even so, the synthetic dimension obtained here is significantly larger than can be obtained with alkali atoms.

Furthermore, a synthetic dimension of even larger sizes, up to hundreds of synthetic lattice sites, can be obtained using essentially the same equipment and approach, if we remove the restriction of using only the $m=0$ or $m=n$ rotational sublevels; unique spectral selectivity of state-to-state transitions is generally still provided by rotational-hyperfine energy shifts at the scale of $\sim 100$~kHz.

\begin{figure}[t]
\begin{tabular}{ccc}
\includegraphics[width=0.3\columnwidth]{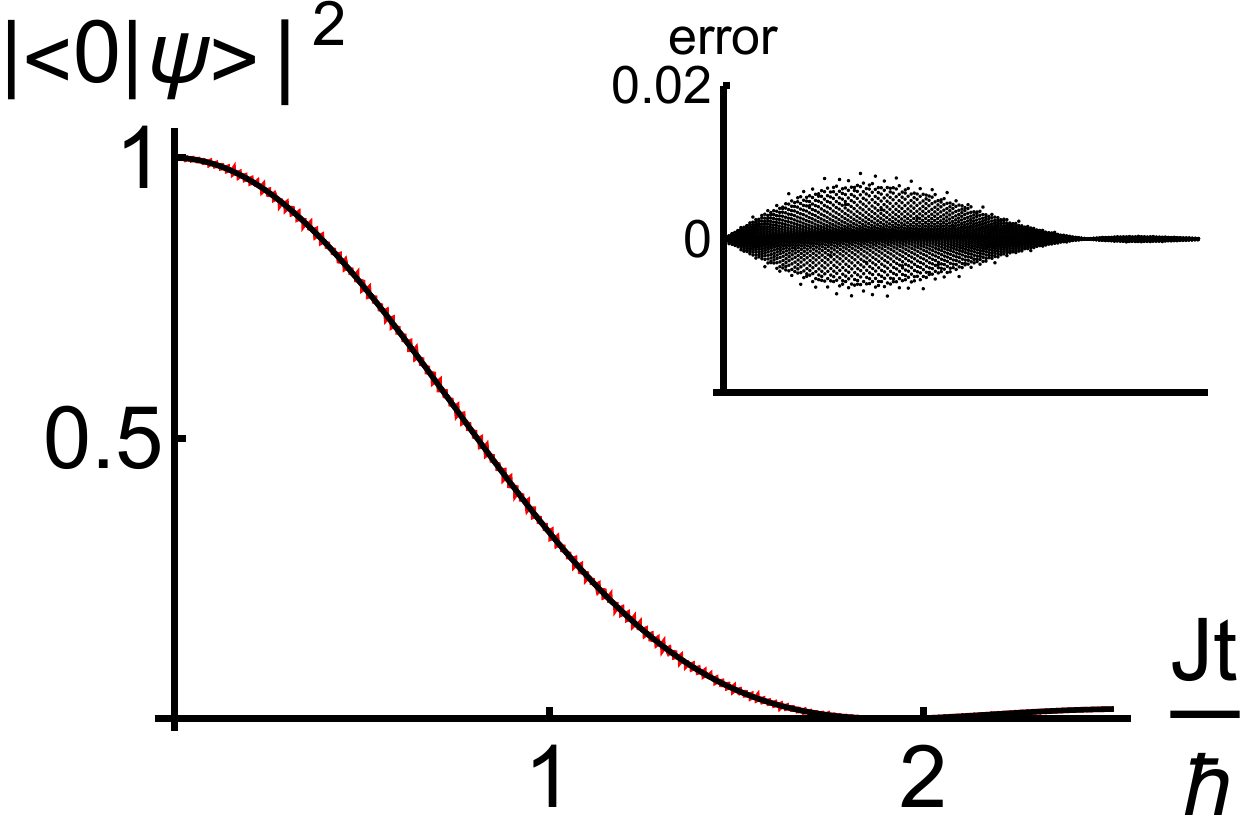} &\qquad & \includegraphics[width=0.3\columnwidth]{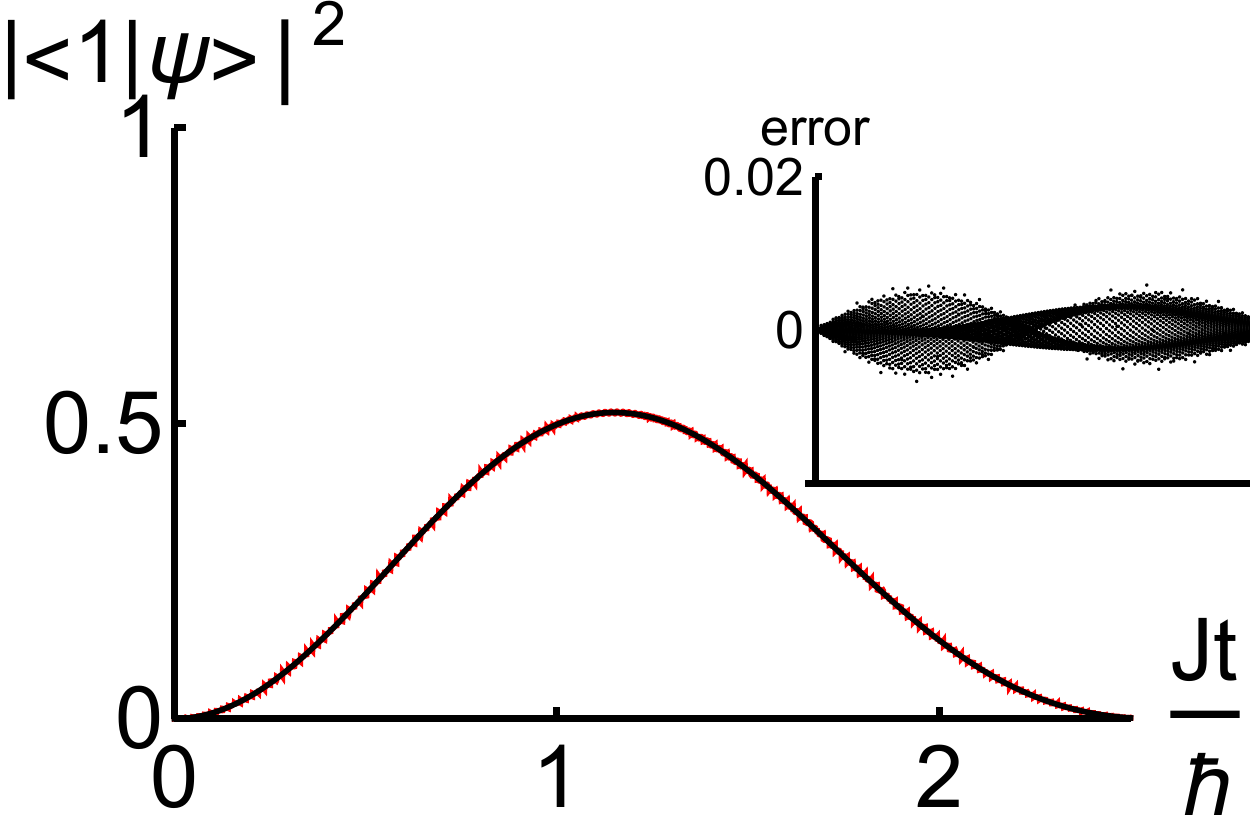}\\
(a) && (b)
\end{tabular}
\caption{The occupation probability of the (a) $\ket{n=0}$ and (b) $\ket{n=1}$ synthetic lattice sites versus time, when the molecule is initialized in $\ket{n=0}$, and evolves under the tight-binding Hamiltonian that is the first term in Eq.~\eqref{eqn: H} of the main text (black) or under the single-particle terms in Eq.~\eqref{eqn: Hsupp} (red) with 100 microwaves. The black and red curves are indistinguishable by eye at the scale shown, demonstrating that the tight binding model accurately describes the molecule's evolution when it is driven by 100 microwaves. The microwave amplitudes were chosen such that $J=B/200$. This is a pessimistically large tunneling; the tight-binding model becomes more accurate as $J/B$ is decreased. We expect that the typical synthetic tunneling will be $J/B\sim10^{-6}$, as explained in the text. Inset: Difference between black and red curves. }
\label{fig: assess RWA}
\end{figure}

The biggest challenge in the implementation of our scheme arises from the fact that different rotational states may experience different real space lattice potentials due to having different ac polarizabilities. This has two consequences: molecules in different rotational states have a) different zero point energies, which shift the microwaves off resonance, and b) different real space Wannier functions, which affect dipole matrix elements. The offsets in the zero point energies can be cancelled by tuning the microwave frequencies appropriately, and/or choosing ``magic angles'' for the polarization of the lattice lasers~\cite{kotochigova2010electric, neyenhuis2012anisotropic}. It may be more practical to tune the frequencies to resonance, because the polarizabilities are very sensitive to small deviations in the polarization angles. The effect on the dipole matrix elements can be compensated by adjusting the respective microwave amplitude until the desired matrix elements for a uniform synthetic tunneling are achieved. The only remaining challenge is then due to inhomogeneities in the real space optical lattice, and a harmonic trap in real space, which will lead to a spatially varying zero point energy. This challenge can be mitigated by using a flat trap, or by observing only a small region of the molecular cloud in a harmonic trap where the optical lattice is relatively uniform in real space. In such a small region, all microwaves can be tuned to resonance with the desired rotational transitions.

One might worry that the illumination of the cloud with a large number of microwaves presents an additional challenge: that the numerous off-resonant driving frequencies lead the molecules to decohere, and that the tight binding model Eq.~\eqref{eqn: H} is no longer valid. However, there is no such decoherence, due to a large separation of energy scales between the microwave frequencies and microwave couplings. For example, a typical microwave frequency used is $\omega_n\sim\mathcal{O}({\rm GHz})$, and a typical synthetic tunneling, for moderate microwave amplitude, is $J\sim\hbar\times\mathcal{O}({\rm kHz})$. For these values, the dynamics of the molecules are modeled accurately by the tight-binding model in Eq.~\eqref{eqn: H}. We illustrate this in Fig.~\ref{fig: assess RWA}, by comparing the dynamics of a molecule governed by [Eq.~\eqref{eqn: H}], and one governed by the single particle terms in Eq.~\eqref{eqn: Hsupp} which include the $100$ microwaves required to create a synthetic dimension with $101$ sites. We find that even for very large synthetic tunnelings $J\sim\frac{\hbar\omega_n}{200}$, the dynamics in the two cases agree to better than $1\%$ accuracy.

\section{Bound state for two molecules}
\begin{figure}[t]
\begin{tabular}{ccc}
\includegraphics[width=0.3\columnwidth]{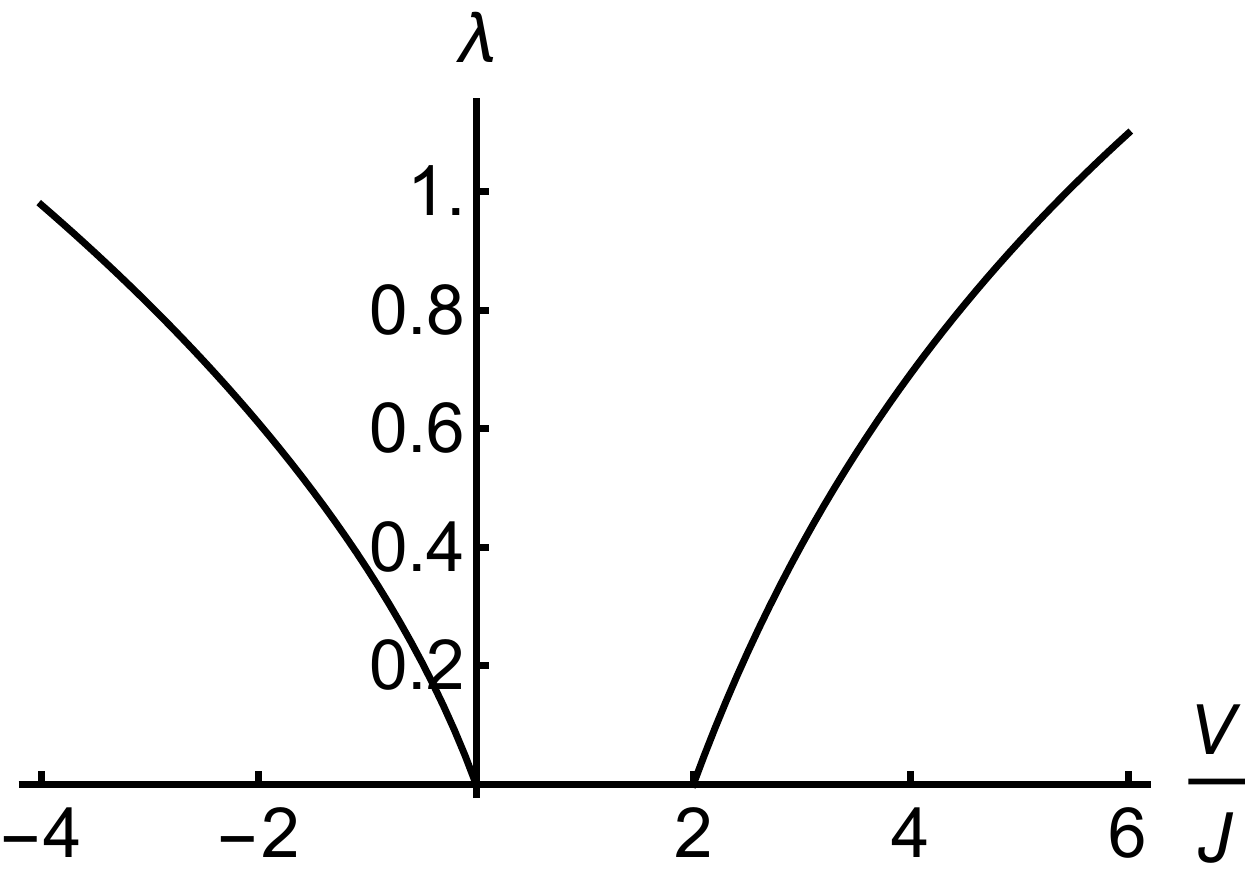} &\qquad & \includegraphics[width=0.3\columnwidth]{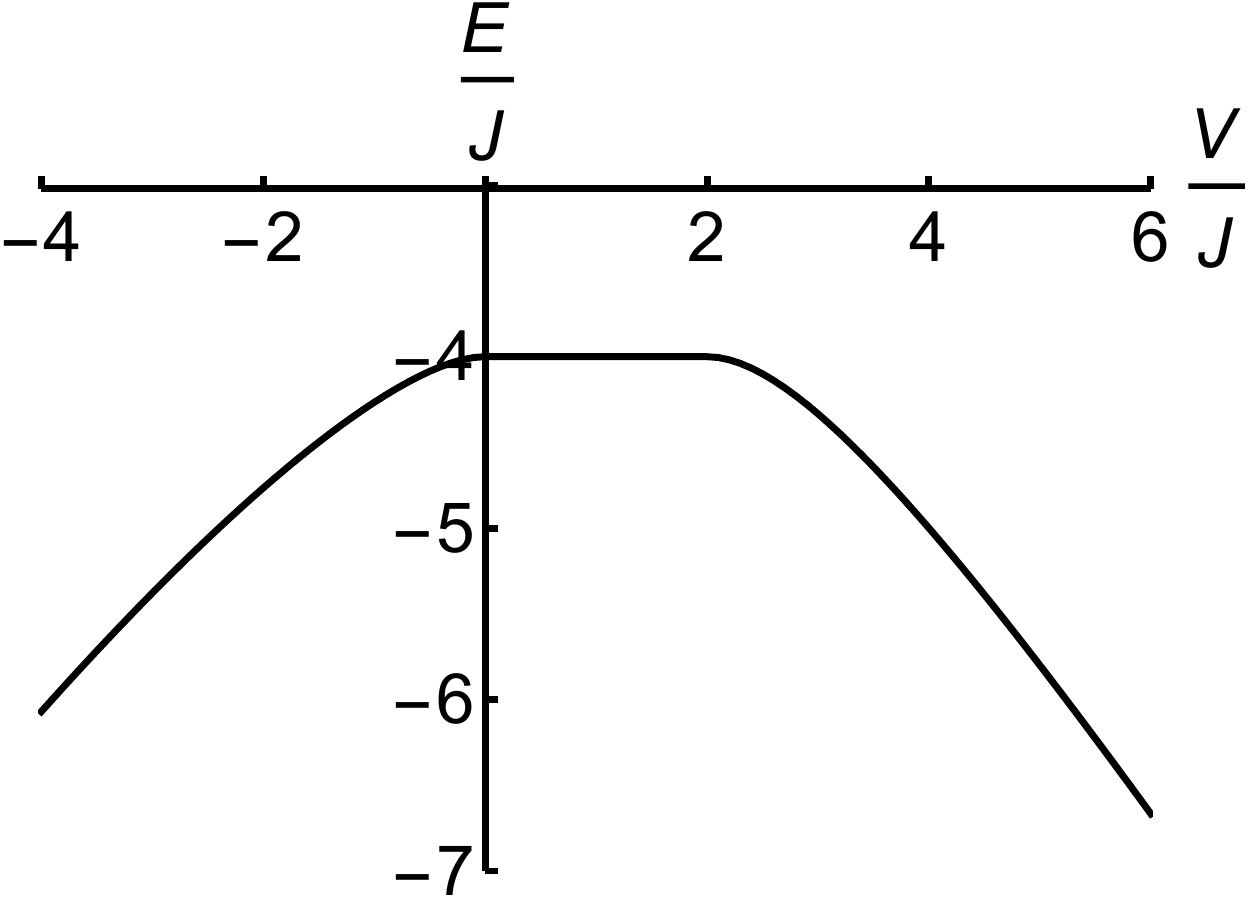}\\
(a) && (b)
\end{tabular}
\caption{(a) The inverse binding length as a function of $V/J$, where $V$ refers to the amplitude for exchanging angular momentum at large $n$. The molecules undergo a transition from an unbound state to a bound state at $V/J=0$ and $V/J=2$. (b) The ground state energy versus $V/J$.}
\label{fig: lambda}
\end{figure}

The Schr\"{o}dinger equation for the relative wavefunction of two molecules, $f(y=m-n)=f_{mn}$, is
\begin{equation}\begin{array}{rcll}
E f(y) &=& -J\left[f(y+1)+f(y-1)\right] &\ {\rm if}\ |y|\neq1\\
E f(y) &=& -J\left[f(y+1)+f(y-1)\right]+Vf(-y) &\ {\rm if}\ |y|=1.
\end{array}\label{eqn: feqns}\end{equation}
We substitute the ansatz that $f(y)=e^{-\lambda y}$ for $y>0$. The first line in Eq. (\ref{eqn: feqns}) tells us that the energy of this bound state is $E=-4J\cosh\lambda$. When $V\leq0$, we find that the ground state satisfies $f(y) = f(-y)$. Using the second line in Eq. (\ref{eqn: feqns}), we find that $\lambda$ solves the implicit transcendental equation:
\begin{equation}
e^\lambda = \frac{2J}{V+4J\cosh\lambda-2J\sech\lambda}.
\label{eqn: trans1}
\end{equation}
The solution for $\lambda$ in Eq. (\ref{eqn: trans1}) is positive for all $V\leq0$. When $V>0$, we find that in the ground state, $f(y)=-f(-y)$. In this case, $\lambda$ solves the transcendental equation:
\begin{equation}
e^\lambda = \frac{-2J}{V-4J\cosh\lambda}.
\label{eqn: trans2}
\end{equation}
The solution for $\lambda$ in Eq. (\ref{eqn: trans2}) is positive only when $V\geq2J$. The positive solution for $\lambda$, and the ground state energy, are plotted in Fig. \ref{fig: lambda}.

\section{Other variational approximations}
In the main text we presented the results for the many-body phase diagram based on a cluster mean field theory. Here we consider two alternative approximations that corroborate the findings there.
\subsection{Single-site mean field approximation}
Our single-site mean field ansatz for a 1D chain of molecules is
\begin{equation}
\ket{\psi'_{\rm var}} = \left(\prod_{i\in{\rm even}}\sum_m g_m \hc_{mi}\+\right)\left(\prod_{j\in{\rm odd}}h_n\hc_{nj}\+\right) \ket{\rm vac}.
\label{eqn: psi1site}
\end{equation}
The variational energy per molecule in this approximation is
\begin{align}
\frac{\expect{\hH}'_{\rm var}}{N_{\rm mol}} = &-\frac{J}{2}\sum_m\left(g_m^*g_{m+1}\nostar + h_m^*h_{m+1}\nostar + \rm{c.c.}\right) + \frac{7\zeta(3)}{8}\sum_{n}V_n^{01}\left(g_n^*g_{n+1}\nostar h_{n+1}^*h_n\nostar + {\rm c.c.}\right)\nonumber\\
 &+ \frac{\zeta(3)}{8}\sum_nV_n^{01}\left(|g_ng_{n+1}|^2+|h_nh_{n+1}|^2\right)
\label{eqn: E1site}
\end{align}
Similar to the cluster mean field approximation in the main text, we found that this ansatz also predicts two binding transitions, which occur at $V\simeq0$ and $V\simeq2.15J$ in the thermodynamic limit. The binding transitions again correspond to a spontaneous dimensional reduction to a fluctuating quantum string. The string hosts an emergent hardcore condensate at $|V/J|=\infty$.

The variational ansatz for a square lattice of molecules is obtained by a straightforward extension of the 1D ansatz in Eq.~\eqref{eqn: psi1site}. That is, bipartition the square lattice into two intercalated checkerboards in the conventional way, run the index $i$ over one checkerboard, and $j$ over the other checkerboard. Here again, we found that two binding transitions occur, similar to the cluster approximation. When $V\simeq0$ or $V\simeq1.5J$, the system undergoes a spontaneous dimensional reduction to a fluctuating quantum membrane. The membrane hosts an emergent hardcore condensate at $|V/J|=\infty$.

\subsection{Fermionic-pair mean field approximation for a 1D chain}
In this approximation, we generalize our system to a grand canonical ensemble of slave fermions, by introducing a chemical potential. The modified Hamiltonian that describes our system is
\begin{equation}
\hH_{\rm eff} =  \sum_j\sum_{n=1}^{N_{\rm rot}} \left(-J_n \left(\hc_{n-1,j}\+\hc_{nj} + {\rm h.c.}\right) - \mu\hc_{nj}\+\hc_{nj}\right) + \sum_{ij}\sum_{n=1}^{N_{\rm rot}} V_n^{ij} \left( \hc_{ni}\+\hc_{n-1,i}\nodag\hc_{n-1,j}\+\hc_{nj}\nodag + {\rm h.c.} \right).
\label{eqn: Hsuppeff2}\end{equation}
We relax the constraint that exactly one molecule occupies each real lattice site, and fix the chemical potential $\mu$ such that this constraint is held on average: $\sum_n\expect{\hc_{nj}\+\hc_{nj}\nodag} = 1\ \forall j$. Motivated by our exact solution for two molecules, we assume mean fields for bound molecules in the synthetic dimension:
\begin{equation}
\Delta_n^{ij} = V_n^{ij}\expect{\hc_{nj}\nodag\hc_{n-1,i}\nodag}.
\end{equation}
To gain a simple understanding, we assume that $V_n^{ij}=V^{ij}$ and $J_n=J>0$ are uniform in $n$. Further, we consider only nearest-neighbor interactions, $V^{ij} = V\delta_{|i-j|,1}$, and periodic boundary conditions in both real and synthetic dimensions. Later we perform a numerical calculation, where we include nonuniform physical values $V_n$ in the low-lying rotational states, and open boundary conditions.

In the fermion-pair approximation, the mean-field Hamiltonian for a 1D chain of molecules is
\begin{equation}
\hH_{\rm MF} = \sum_{nj} \left(-J\left(\hc_{nj}\+\hc_{n-1,j} + {\rm h.c.}\right) -\mu\hc_{nj}\+\hc_{nj}\nodag + \left(\Delta\hc_{nj}\+\hc_{n-1,j+1}\+ + \Delta'\hc_{n,j+1}\+\hc_{n-1,j}\+ + {\rm h.c.}\right)\right). \label{eqn: HsuppMF}
\end{equation}
Here, $\Delta=\Delta_n^{j,j+1}$ and $\Delta'=\Delta_n^{j+1,j}$. We self-consistently diagonalize $\hH_{\rm MF}$ to find $\Delta,\ \Delta'$ and $\mu$. We find different solutions when $V<0$ and $V>0$. In both regimes, the parameters $\Delta,\ \Delta'$ and $\mu$ solve the implicit equations:
\begin{align}
\frac{1}{|V|}\ =\ & \frac{2}{N_{\rm rot}N_{\rm mol}}\sum_{\vk} \frac{ \left|g^2(\vk)\right| }{\sqrt{\left(\mu+2J\cos k_Z\right)^2 + 16\Delta^2\left|g^2(\vk)\right|}},\\
1\ =\ & \frac{1}{2N_{\rm mol}}\sum_{\vk} \left(1+\frac{ \mu+2J\cos k_Z}{\sqrt{\left(\mu+2J\cos k_Z\right)^2 + 16\Delta^2\left|g^2(\vk)\right|}}\right).
\end{align}
When $V<0$, $\Delta=\Delta'$ and $g(\vk) = \cos k_Z\sin k_X$ is a self-consistent solution. When $V>0$, $\Delta=-\Delta'$ and $g(\vk) = \sin k_Z\cos k_X$  is a self-consistent solution. The free energy per molecule is
\begin{equation}
\frac{\expect{\hH}_{\rm MF}}{N_{\rm mol}} = -\frac{1}{2N_{\rm mol}}\sum_{\vk} \left(\mu+2J\cos k_Z - \frac{4|\Delta|^2}{|V|} + \sqrt{\left(\mu+2J\cos k_Z\right)^2 + 16\Delta^2g^2(\vk)}\right).
\label{eqn: EMF}
\end{equation}
Here, $X$ and $Z$ refer to the real and synthetic directions (not the same as $x$ and $z$), and $k_X$ and $k_Z$ are momenta conjugate to $X$ and $Z$. The mean field ground states are achiral $p$-wave superfluids of slave fermions. The order parameter is even (odd) along the synthetic direction, and odd (even) along the real direction, when $V<0$ ($V>0$). This is related to the parity of the two-molecule ground state when $V<0$ and $V>0$. In the thermodynamic limit of $N_{\rm rot}$ and $N_{\rm mol}$, the system is in the superfluid phase for $V<0$ and $V>2J$. For $N_{\rm rot}\gg1$ and $V/J\rightarrow\pm\infty$, the mean field superfluids are identical to the two-molecule ground state in Eq.~\eqref{eqn: psi2mol} when they are projected onto the physical sector of unit molecular filling. The system undergoes a phase transition to a normal Fermi liquid phase at $V=0$ and $V=2J$.

\subsection{Fermionic-pair mean field approximation in 2D: chiral superfluids}
\begin{figure}[t]
\includegraphics[width=0.3\columnwidth]{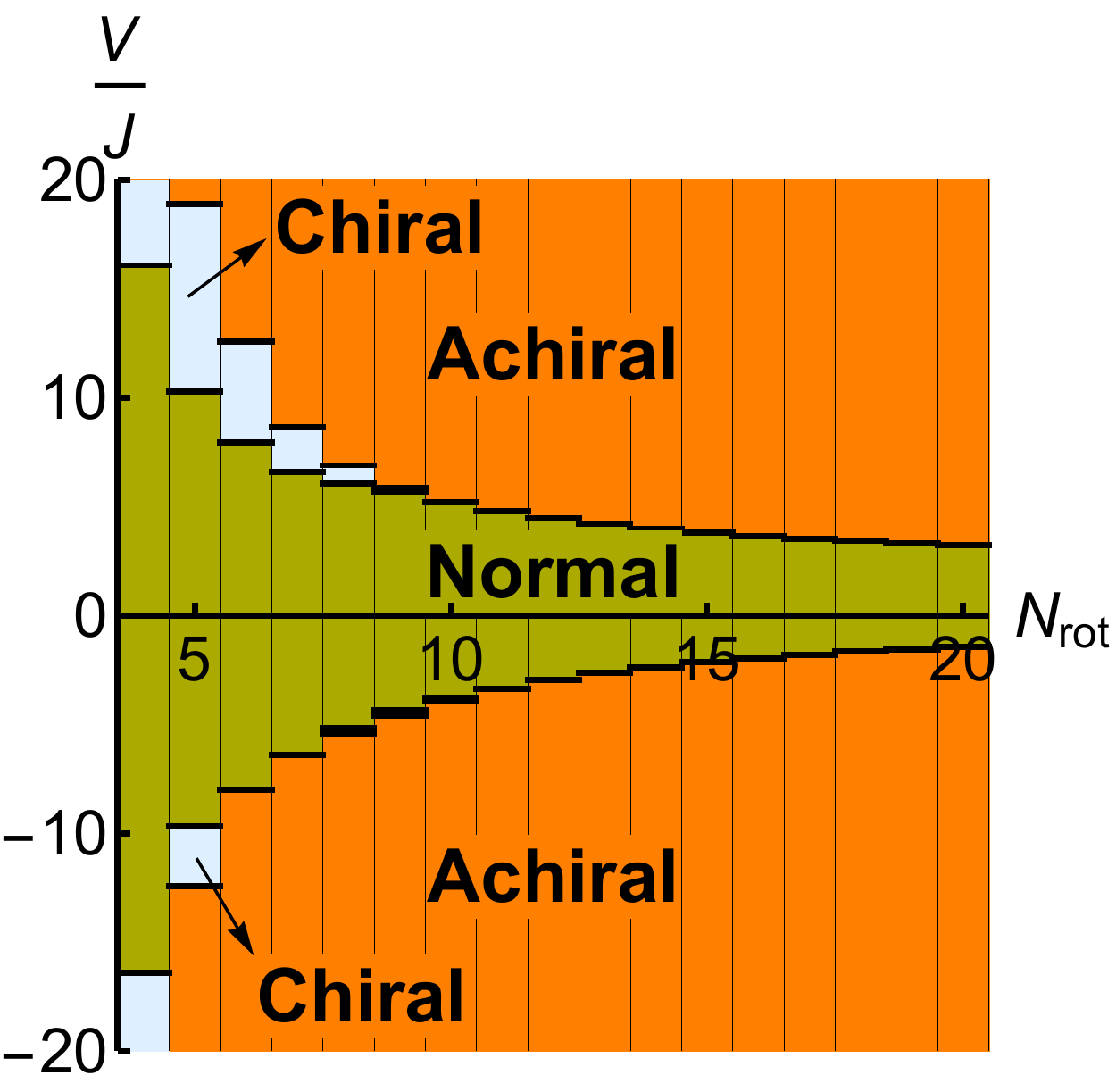}
\caption{Phase diagram of a 2D array of molecules under a fermionic-pair mean field approximation. Chiral $p_x+ip_y$ superfluids emerge in a narrow region of $V/J$ for small $N_{\rm rot}$.}
\label{fig: chiralAchiralSF}
\end{figure}
In a 2D array of molecules, the fermionic-pair mean field approximation results in chiral as well as achiral superfluid phases. The mean-field Hamiltonian for a 2D array, assuming only nearest-neighbor dipole interactions $V_n^{ij}=V\delta_{|i-j|=1}$, is
\begin{align}
\hH_{\rm MF} = \sum_{nj} &\left(-J\left(\hc_{nj}\+\hc_{n-1,j} + {\rm h.c.}\right) -\mu\hc_{nj}\+\hc_{nj}\right.\nonumber\\
& \left.+ \left(\Delta_X\hc_{nj}\+\hc_{n-1,j+\hat{X}}\+ + \Delta_X'\hc_{n,j+\hat{X}}\+\hc_{n-1,j}\+ + \Delta_Y\hc_{nj}\+\hc_{n-1,j+\hat{Y}}\+ + \Delta_Y'\hc_{n,j+\hat{Y}}\+\hc_{n-1,j}\+ + {\rm h.c.}\right)\right),
\end{align}
where $\Delta_{A}=\Delta_n^{j,j+\hat{A}}$ and $\Delta_A'=\Delta_n^{j+\hat{A},j}$ ($\hat{A}=\hat{X}\ \rm{or}\ \hat{Y}$). When the dipole interactions $V_n$ are equal and negative in both real directions, there are two allowed self-consistent solutions:
\begin{equation}\begin{array}{lcl}
\Delta_X = \Delta_X'= \Delta_Y = \Delta_Y', && g(\vk) = \cos k_Z\left(\sin k_X+\sin k_Y\right),\ {\rm or}\\
\Delta_X = \Delta_X'= i\Delta_Y = i\Delta_Y', && g(\vk) = \cos k_Z\left(\sin k_X+i\sin k_Y\right).
\end{array}\label{eqn: 2DMF1}\end{equation}
The first solution in Eq. (\ref{eqn: 2DMF1}) describes an achiral superfluid, while the second solution is chiral. When the dipole interactions $V_n$ are equal and positive in both real directions, there are two allowed self-consistent solutions:
\begin{equation}\begin{array}{lcl}
\Delta_X = -\Delta_X'= \Delta_Y = -\Delta_Y', && g(\vk) = \sin k_Z\left(\cos k_X+\cos k_Y\right),\ {\rm or}\\
\Delta_X = -\Delta_X'= i\Delta_Y = -i\Delta_Y', && g(\vk) = \sin k_Z\left(\cos k_X+i\cos k_Y\right).
\end{array}\label{eqn: 2DMF2}\end{equation}
The first solution again describes an achiral superfluid, while the second solution is chiral.

We calculate the energy of the chiral and achiral superfluids from Eq.~\eqref{eqn: EMF}, and plot the superfluid phase with the least energy in Fig.~\ref{fig: chiralAchiralSF}. We find that the chiral phase is the most stable superfluid in a narrow region of $V/J$ for small $N_{\rm rot}$. When projected to the physical subspace, the chiral superfluids correspond to non-Abelian Ising anyonic phases~\cite{burnell20112}.

\subsection{Comparison of energy in various approximations}\label{subsec: MFcomparison}
\begin{figure}[t]
\includegraphics[width=0.3\columnwidth]{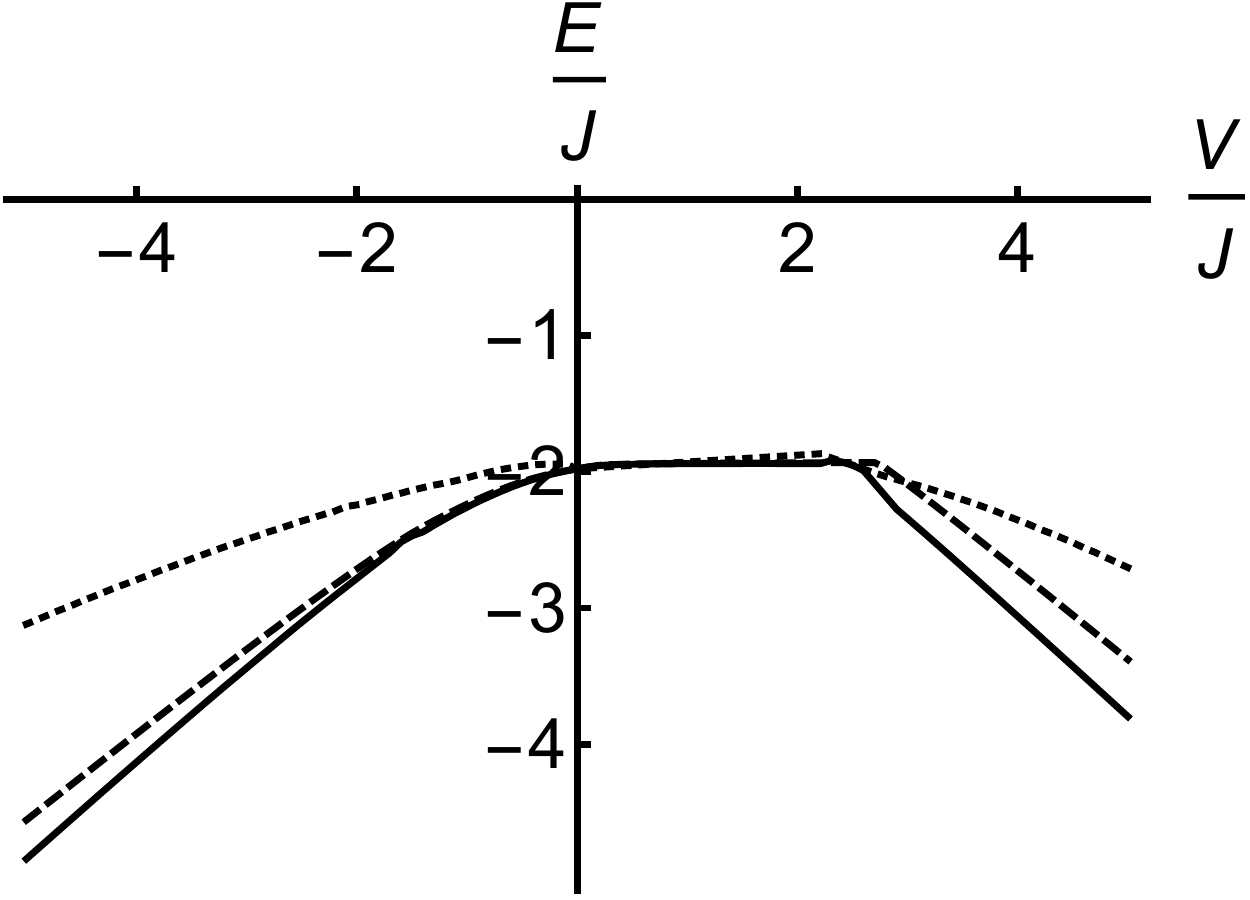}
\caption{The ground state energy for a 1D chain of molecules in three different approximations: a cluster mean field approximation (solid), single-site mean field (dashed), and fermion-pair mean field (dotted). The calculation was performed with only nearest-neighbor dipolar interactions and the physical values of $V_n$. The size of the synthetic dimension was chosen to be $N_{\rm rot}=20$. The cluster mean field state always has the lowest energy. The energies are similar when the full long-ranged dipolar interaction is included.}
\label{fig: comparison}
\end{figure}
In a 1D chain, the total variational energy per molecule in the cluster mean field approximation [Eq. (\ref{eqn: psivar})] is
\begin{align}
\frac{\expect{\hH}_{\rm var}}{N_{\rm mol}} = &\left(-\frac{J}{2}\sum_{mn}f^*_{mn}\left(f_{m,n+1}\nostar+f_{m+1,n}\nostar\right) + \frac{1}{2}\sum_n V_n^{01} f_{n,n+1}^*f_{n+1,n}\nostar + \frac{7\zeta(3)-4}{8}\sum_{mm'n}V_n^{01}f_{mn}^*f_{n+1,m'}^*f_{m,n+1}\nostar f_{n,m'}\nostar\right. \nonumber \\
&\left.+ \frac{\zeta(3)}{16}\sum_{mm'n} V_n^{01}\left(f_{mn}^*f_{m',n+1}^*f_{m,n+1}\nostar f_{m',n}\nostar + f_{nm}^*f_{n+1,m'}^*f_{n+1,m}\nostar f_{n,m'}\nostar\right)\right) + {\rm c.c}.
\label{eqn: E2site}
\end{align}
The 1D variational energy in the single-site and fermionic-pair mean field approximations were given in Eqs.~\eqref{eqn: E1site} and~\eqref{eqn: EMF}. For simplicity, in this section, we modify the expressions to include only nearest-neighbor dipole interactions, and plot the variational minimum in the three approximations in Fig.~\ref{fig: comparison}.The physics is unchanged when the full dipole interaction is included. We find that the cluster mean field state always has the lowest energy. The single-site mean field state has nearly the same energy as the cluster mean field state for $V<2J$, and the fermionic-pair state displays similar behavior as the single-site and cluster mean field states. These observations suggest that all three approximations capture the essential features of the ground state, and the cluster mean field state is the closest approximation to the true ground state.

We also calculated the variational energies of the ground state of a square lattice of molecules in the three approximations. We again found that the cluster mean field state has the lowest energy, and the other approximations result in similar energies.
\end{widetext}
\end{document}